\documentclass[11pt,a4paper]{article}

\usepackage[T1]{fontenc}
\usepackage[utf8]{inputenc}

\usepackage{amsmath,amssymb,amsfonts}
\usepackage{graphicx}
\usepackage{booktabs}
\usepackage{multirow}
\usepackage{hyperref}
\usepackage{physics}
\usepackage{geometry}
\usepackage{subcaption}
\usepackage{mathtools}
\usepackage{pdflscape}
\usepackage{xcolor}
\usepackage{cite}
\usepackage{hyperref}

\title{\bfseries Bound, Scattering, and Resonance States \\ of the Symmetric Woods--Saxon Potential \\in Dunkl Quantum Mechanics

}

\author{
Mebarek Heddar$\,^{1}$, %ORCID: 0000-0002-7217-1832,
Can Ertu\u{g}ay$\,^{2}$, %ORCID: 0000-0001-9508-0207
Bekir Can L\"utf\"uo\u{g}lu$^{\,3,\dagger}$ %ORCID: 0000-0001-6467-5005
}

\date{}

\begin{document}

\maketitle

\begin{center}
\small

$^{1}$ Laboratory of Photonic Physics and Nano-Materials (LPPNNM),\\
Department of Matter Sciences, University of Biskra, Algeria

\vspace{0.4em}

$^{2}$ Department of Physics, Faculty of Science, Akdeniz University, Antalya, Turkey

\vspace{0.4em}

$^{3}$ Department of Physics, Faculty of Science, University of Hradec Kr\'alov\'e,\\
Rokitansk\'eho 62/26, 500\,03 Hradec Kr\'alov\'e, Czech Republic

\vspace{0.8em}

$^{\dagger}$Corresponding author:\\
\href{mailto:bekir.lutfuoglu@uhk.cz}{\texttt{bekir.lutfuoglu@uhk.cz}}

\end{center}

\begin{abstract}
We present the first unified analytical description of the bound, scattering, and Gamow (metastable) resonance states of the one-dimensional symmetric Woods--Saxon potential within the framework of Dunkl quantum mechanics, establishing a single analytical framework that consistently describes all physically relevant spectral regimes. The Dunkl deformation introduces a reflection-sector-dependent inverse-square interaction that precludes a direct analytical treatment of the corresponding Schr\"odinger equation. By employing a Pekeris-type approximation, the problem is transformed into a hypergeometric differential equation, enabling the derivation of analytical wave functions, bound-state quantization conditions, scattering amplitudes, and Gamow resonance solutions within a unified formalism. The results reveal that the Dunkl deformation fundamentally modifies the system's spectral properties by generating a reflection-sector-dependent splitting of the bound-state spectrum, along with quantitative changes in the reflection and transmission probabilities. The narrow transmission peaks are shown to originate from long-lived Gamow resonance states, whose complex energies, decay widths, and lifetimes are determined through analytic continuation into the complex-energy plane. In the limit of vanishing Dunkl deformation, the conventional Woods--Saxon results are fully recovered, confirming the consistency of the proposed formalism. The proposed analytical framework provides a versatile approach for investigating finite-range potentials in Dunkl quantum mechanics and opens the way to systematic studies of reflection-deformed quantum systems beyond the Woods--Saxon interaction.
\end{abstract}

\section{Introduction}

Reflection symmetry plays a fundamental role in many areas of mathematical physics and quantum mechanics. Quantum systems possessing discrete reflection symmetry often admit algebraic structures that extend the conventional Heisenberg algebra through the incorporation of reflection operators. Such generalized frameworks have found applications in generalized statistics, supersymmetric quantum mechanics, exactly solvable models, and quantum systems with deformation-induced interactions.

One of the earliest systematic investigations in this direction was carried out by Wigner, who questioned whether the canonical commutation relations are uniquely determined by the equations of motion. By analyzing the free particle and the harmonic oscillator, he demonstrated that the equations of motion alone do not uniquely determine the canonical commutation relations, thereby opening the possibility of alternative algebraic structures compatible with the same quantum dynamics~\cite{Wigner:1950}. This ambiguity was subsequently addressed by Yang, who reconsidered Wigner's generalized oscillator within a rigorous Hilbert-space formulation. Remarkably, the corresponding operator representation naturally involves a reflection operator, establishing one of the earliest links between generalized quantum algebras and reflection symmetry~\cite{Yang:1951}.  Green further generalized these ideas by introducing a field quantization scheme based on trilinear commutation relations, thereby establishing the theory of parabose and parafermi statistics as an extension of the conventional Bose--Einstein and Fermi--Dirac statistics \cite{Green:1952kp}. The physical implications of generalized statistics were later investigated by Greenberg, who demonstrated that particles obeying parastatistics can be consistently incorporated into quantum field theory \cite{Greenberg:1964pe}.

Remarkably, a conceptually similar operator emerged several decades later in a completely different mathematical context. Motivated by harmonic analysis and the theory of finite reflection groups, Dunkl introduced a family of differential--difference operators associated with finite reflection groups, providing a unified mathematical framework that naturally combines differential and reflection operators~\cite{Dunkl:1989dvp}. These operators preserve many of the analytical properties of ordinary differential operators while encoding the underlying reflection symmetry of the system. Their mathematical structure was further developed through the introduction of the Dunkl kernel and related analytical constructions, establishing the foundations of modern Dunkl analysis~\cite{Dunkl:1991}. Owing to their intrinsic connection with reflection symmetry, Dunkl operators are also closely related to reflection-deformed Heisenberg algebras and other generalized quantum algebras involving reflection operators. In particular, the pioneering studies of Plyushchay and collaborators established reflection-deformed Heisenberg algebras as a versatile framework for generalized oscillator systems, parabosonic models, bosonized and hidden supersymmetry, and particles with exotic spin and statistics. These studies further highlighted the fundamental role of reflection operators in generalized quantum algebras and exactly solvable quantum systems~\cite{Plyushchay:1993ak,Plyushchay:1994re,Plyushchay:1996zlq,Plyushchay:1997ty,Plyushchay:1997ie,Klishevich:1999nq,Gamboa:1998bz,Plyushchay:1999qz,Klishevich:2001gy,Horvathy:2010vm}.

Building upon these mathematical developments, Dunkl operators were systematically incorporated into quantum mechanics through the study of exactly solvable and superintegrable systems. Early investigations focused on the Dunkl harmonic oscillator, which soon became a prototype model for exploring the role of reflection symmetry in quantum mechanics. These studies led to the construction of the Schwinger--Dunkl algebra, the realization of the Bannai--Ito algebra as the corresponding symmetry algebra, and the development of a variety of superintegrable quantum systems, thereby establishing the foundations of Dunkl quantum mechanics and its algebraic structure~\cite{Genest:2012exa,Genest:2013spj,Genest:2014a,Genest:2014b,Genest:2015,DeBie:2015,DeBie:2016}.

Subsequent developments extended these algebraic foundations to a broad range of physical applications. Analytical investigations initially focused on the Dunkl harmonic oscillator and related exactly solvable systems, establishing the first nontrivial applications of the formalism in quantum mechanics~\cite{Salazar-Ramirez:2016bpi,Salazar-Ramirez:2016mfj,Ghazouani:2019,Sedaghatnia:2023rvh,Quesne:2023rgc}. The formalism was subsequently generalized to the Schr\"odinger equation with various interaction potentials~\cite{Chung:2019eok,Mota:2021tye,Schulze-Halberg:2022rsn,Arabsaghari:2024xxs,Hamil:2024feo}, followed by relativistic extensions to the Dirac, Klein--Gordon, and Duffin--Kemmer--Petiau equations~\cite{Sargolzaeipor:2018rfl,Mota:2018qgn,Ojeda-Guillen:2020mul,Merad:2021mqq,Hamil:2021dhq,Rouabhia:2023mmo,Hamil:2024iau,Askari:2023kzo}. More recently, the scope of Dunkl quantum mechanics has expanded further to include higher-dimensional systems, path-integral formulations, generalized Dunkl operators, finite-range interaction potentials, and noncommutative quantum mechanics~\cite{Benchikha:2024ebc,Benchikha:2024qhm,Benchikha:2024kov,Schulze-Halberg:2024gij,Schulze-Halberg:2024nmw,Schulze-Halberg:2024lpy,Benzair:2025iee,Benzair:2025czn}. Applications to quantum statistical mechanics and thermodynamic systems have also continued to grow, including studies of Bose gases, blackbody radiation, and related many-body phenomena~\cite{Dong:2021,Hassanabadi:2021swd,Hamil:2022uwy,Merabtine:2023cso,Hamil:2023avs,Hocine:2023foq,Hocine:2023ybl,Benarous:2024cca}.

Among finite-range interaction potentials, the Woods--Saxon (WS) potential occupies a central position in quantum physics. Originally introduced as a three-dimensional radial potential to describe the nuclear mean field, it provides a realistic representation of finite nuclei through its finite depth and diffuse surface~\cite{Woods:1954zz,BohrMottelson}. Owing to these properties, the WS potential and its one-dimensional symmetric form (sWS) have been widely employed in studies of bound states, scattering, tunneling, and resonance phenomena using a variety of analytical and numerical approaches~\cite{ Kennedy:2002,Berkdemir:2005vz,Tezcan:2008ylp,Bayrak:2014hda,Lutfuoglu:2016red}. While the conventional WS potential has been extensively studied within ordinary quantum mechanics, its extension to Dunkl quantum mechanics remains relatively unexplored, particularly with regard to the unified description of bound, scattering, and resonance states.

This work provides the first unified analytical treatment of the bound, scattering, and Gamow (metastable) resonance states of the one-dimensional sWS potential within Dunkl quantum mechanics. By employing a Pekeris-type approximation, the reflection-dependent inverse-square interaction generated by the Dunkl deformation is incorporated into the effective potential, allowing the corresponding Schr\"odinger equation to be reduced to a hypergeometric form. Analytical expressions for the wave functions, bound-state quantization conditions, scattering amplitudes, and Gamow resonance solutions are derived within a common formalism. Particular attention is devoted to the influence of the Dunkl deformation on the bound-state spectrum, reflection and transmission coefficients, resonance energies, decay widths, and lifetimes, while the undeformed limit is shown to recover the conventional WS results.

The remainder of this paper is organized as follows. Section \ref{Sec:sec2} presents the one-dimensional Dunkl--Schr\"odinger equation, the effective symmetric Woods--Saxon potential, and the analytical framework adopted in this work. Sections \ref{Sec:sec3} and \ref{Sec:sec4}  are devoted to the bound-state and scattering-state solutions, respectively, with the latter including the investigation of Gamow (metastable) resonance states. Finally, the main conclusions are presented in Section \ref{Sec:sec5}.

\section{Theoretical Framework}
\label{Sec:sec2}

In this section, we formulate the one-dimensional Dunkl--Schr\"odinger equation for a particle interacting through the Woods--Saxon potential and derive the corresponding effective potential. Owing to the reflection symmetry inherent in the Dunkl formalism, the Hilbert space decomposes into two independent reflection sectors. This property enables a unified analytical treatment of both the scattering and bound-state problems presented in the following sections.

\subsection{One-Dimensional Dunkl Schr\"odinger Equation}
\label{Sec:subsec21}

The one-dimensional Dunkl derivative operator is defined as \cite{Dunkl:1989dvp}
\begin{equation}
\hat{D}_{x}
=
\frac{d}{dx}
+
\frac{\mu}{x}
\left(
1-\hat{R}
\right),
\label{eq:DunklDerivative}
\end{equation}
where $\mu>-1/2$ is the Dunkl deformation parameter and $\hat{R}$ denotes the reflection operator acting on a sufficiently smooth function according to
\begin{equation}
\hat{R}f(x)=f(-x).
\label{eq:ReflectionOperator}
\end{equation}

The Dunkl derivative combines differential and reflection operations within a single operator, thereby providing a natural framework for incorporating discrete reflection symmetry into quantum mechanics. The reflection operator satisfies the algebraic relations
\begin{equation}
\hat{R}x=-x\hat{R},
\qquad
\hat{R}\frac{d}{dx}
=
-\frac{d}{dx}\hat{R},
\label{eq:ReflectionRelations}
\end{equation}
which immediately implies
\begin{equation}
\hat{R}\hat{D}_{x}
=
-\hat{D}_{x}\hat{R},
\qquad
[\hat{R},\hat{D}_{x}^{\,2}]=0.
\label{eq:DunklRelations}
\end{equation}
Using the algebraic relations given above, the squared Dunkl derivative can be written as
\begin{equation}
\hat{D}_{x}^{\,2}
=
\frac{d^{2}}{dx^{2}}
+
\frac{2\mu}{x}\frac{d}{dx}
-
\frac{\mu}{x^{2}}
\left(
1-\hat{R}
\right),
\label{eq:DunklDerivativeSquare}
\end{equation}
which explicitly exhibits the reflection-dependent contribution introduced by the Dunkl deformation.

The one-dimensional Dunkl Hamiltonian is therefore given by
\begin{equation}
\hat{H}
=
-\frac{\hbar^{2}}{2m}
\hat{D}_{x}^{2}
+
V(x),
\label{eq:DunklHamiltonian}
\end{equation}
where $V(x)$ denotes the interaction potential. Throughout this work, we restrict our analysis to reflection-symmetric potentials satisfying
\begin{equation}
V(x)=V(-x).
\label{eq:SymmetricPotential}
\end{equation}

For reflection-symmetric potentials, Eq.~(\ref{eq:DunklRelations}) ensures that the Hamiltonian commutes with the reflection operator,
\begin{equation}
[\hat{H},\hat{R}]=0,
\label{eq:CommutationRelation}
\end{equation}
which allows the eigenfunctions of the Hamiltonian to be chosen simultaneously as eigenfunctions of $\hat{R}$,
\begin{equation}
\hat{R}\psi^{(\omega)}(x)
=
\omega\,\psi^{(\omega)}(x),
\qquad
\omega=\pm1,
\label{eq:ReflectionEigenvalue}
\end{equation}
where $\omega$ labels the reflection sector.

The reflection sectors characterized by the eigenvalues $\omega=\pm1$ should not be confused with the spatial parity of the wave function. The quantity $\omega$ specifies the eigenvalue of the Dunkl reflection operator, whereas the spatial parity is determined by the symmetry of the wave function under the transformation $x\rightarrow -x$. Consequently, each reflection sector may, in principle, accommodate both even- and odd-parity solutions depending on the boundary conditions imposed on the physical problem.

Using Eqs.~(\ref{eq:DunklHamiltonian}), (\ref{eq:ReflectionEigenvalue}), and (\ref{eq:DunklDerivativeSquare}),  one obtains the one-dimensional stationary Dunkl--Schr\"odinger equation for a reflection-symmetric potential
\begin{equation}
\frac{d^2 \psi^{(\omega)}(x)}{dx^2}
+
\frac{2\mu}{x}\frac{d \psi^{(\omega)}(x)}{dx}
+
\left[
\frac{2m}{\hbar^{2}}
\Big(
E-V(x)
\Big)
-
\frac{\mu(1-\omega)}{x^{2}}
\right]
\psi^{(\omega)}(x)
=
0.
\label{eq:DunklSchrodinger}
\end{equation}
The Dunkl deformation introduces an additional first-derivative term together with a reflection-dependent inverse-square contribution. Unlike the first-derivative term, the latter explicitly depends on the reflection eigenvalue $\omega$, causing the two reflection sectors to exhibit distinct quantum dynamics and, consequently, different bound-state spectra and scattering characteristics. In the limit $\mu\rightarrow0$, both Dunkl-induced terms vanish, and Eq.~(\ref{eq:DunklSchrodinger}) reduces to the conventional one-dimensional Schr\"odinger equation.

In the present work, the interaction between the particle and the external field is described by the one-dimensional sWS potential
\begin{equation}
V(x)=
\frac{\sigma V_{0}}
{1+e^{a(|x|-L)}},
\label{eq:WSPotential}
\end{equation}
where $V_{0}$ denotes the potential strength, $L$ determines the width of the potential, and $a$ controls the steepness of the potential surface. The parameter $\sigma$ distinguishes the barrier and well configurations according to
\begin{equation}
\sigma=
\begin{cases}
+1, & \text{barrier potential},\\
-1, & \text{well potential}.
\end{cases}
\label{eq:SigmaDefinition}
\end{equation}

To eliminate the first-derivative term appearing in Eq.~(\ref{eq:DunklSchrodinger}), we introduce the transformation

\begin{equation}
\psi^{(\omega)}(x)=\abs{x}^{-\mu}\phi^{(\omega)}(x).
\label{eq:WaveTransformation}
\end{equation}
Substituting Eq.~(\ref{eq:WaveTransformation}) into Eq.~(\ref{eq:DunklSchrodinger}) yields a Schr\"odinger-type equation of the form
\begin{equation}
\frac{d^2\phi^{(\omega)}(x)}{dx^2} + \frac{2m}{\hbar^2} \Big[E - V^{(\omega)}_{\mathrm{eff}}(x) \Big] \phi^{(\omega)}(x) = 0,
\label{eq:EffectiveSchrodinger0}
\end{equation}
where
\begin{equation}
V^{(\omega)}_{\mathrm{eff}}(x) = \frac{\sigma V_{0}}{1+e^{a(|x|-L)}} + \frac{\hbar^2}{2m} \frac{\mu(\mu-\omega)} {x^2}.\label{eq:FullEffectivePotential}
\end{equation}
The effective potential consists of the one-dimensional sWS potential supplemented by an inverse-square term generated by the Dunkl deformation. Its strength is determined by the eigenvalue $\omega$ of the reflection operator $\hat{R}$, yielding the coefficients $\mu(\mu-1)$ and $\mu(\mu+1)$ for $\omega=+1$ and $\omega=-1$, respectively. Consequently, the two sectors experience different effective potentials. This inverse-square contribution is not merely a correction to the sWS potential, but an intrinsic structural feature of the Dunkl formalism that influences both the scattering characteristics and the bound-state spectrum.

%It should be emphasized that the label $\omega$ identifies the eigenvalue of the Dunkl reflection operator rather than the spatial parity of the wave function. Consequently, each sector may contain both even- and odd-parity solutions, depending on the imposed boundary conditions. 

The inverse-square contribution appearing in Eq.~(\ref{eq:FullEffectivePotential}) prevents a direct reduction of the effective Schr\"odinger equation into a hypergeometric form. To overcome this difficulty, we employ a Pekeris-type approximation in the vicinity of potential boundaries. This approximation allows the centrifugal-like term to be expressed in terms of the same functional structure as the WS potential, thereby enabling an approximate analytical treatment of the problem.

Expanding the inverse-square term about the points $x=\pm L$, the Pekeris-type approximation takes the form
\begin{equation}
\frac{1}{x^2} \approx \frac{1}{L^2} \left[ C_0 + \frac{C_1}{1+e^{a(|x|-L)}} + \frac{C_2}{\left(1+e^{a(|x|-L)}\right)^2} \right], \label{eq:PekerisApproximation}
\end{equation}
where the dimensionless coefficients are
\begin{equation}
C_0 = 1-\frac{4}{\alpha} +\frac{12}{\alpha^2}, \qquad
C_1 = \frac{8}{\alpha} - \frac{48}{\alpha^2},
\qquad
C_2 = \frac{48}{\alpha^2}, \label{eq:PekerisCoefficients}
\end{equation}
with
\begin{equation}
\alpha=aL. \label{eq:AlphaDefinition}
\end{equation}
Substituting Eq.~(\ref{eq:PekerisApproximation}) into Eq.~(\ref{eq:FullEffectivePotential}) yields the approximate effective potential 
\begin{equation}
V^{(\omega)}_{\mathrm{eff}}(x) \approx \frac{\sigma V_0}{1+e^{a(|x|-L)}} + \delta^{(\omega)} \left[ C_0 + \frac{C_1}{1+e^{a(|x|-L)}} + \frac{C_2}{\left(1+e^{a(|x|-L)}\right)^2} \right], \label{eq:ApproximatePotential}
\end{equation}
where
\begin{equation}
\delta^{(\omega)} = \frac{\hbar^2}{2m} \frac{\mu(\mu-\omega)}{L^2}. \label{eq:DeltaDefinition}
\end{equation}
For the sake of simplicity, the superscript $(\omega)$ indicating the reflection sector will be omitted from the notation hereafter.

\begin{figure}[htb!]
\centering

\begin{subfigure}{0.48\textwidth}
\centering
\includegraphics[width=\linewidth]{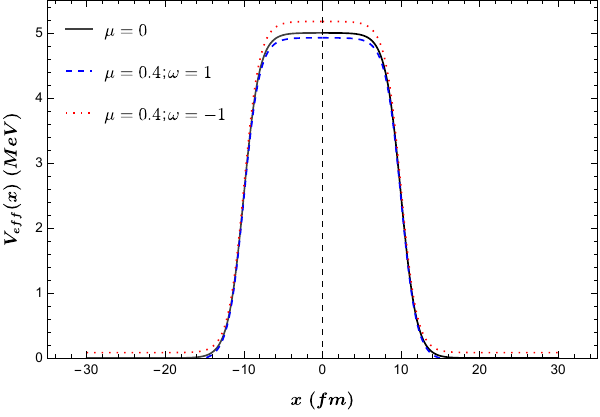}
\caption{Barrier configuration ($\sigma=+1$).}
\end{subfigure}
\hfill
\begin{subfigure}{0.48\textwidth}
\centering
\includegraphics[width=\linewidth]{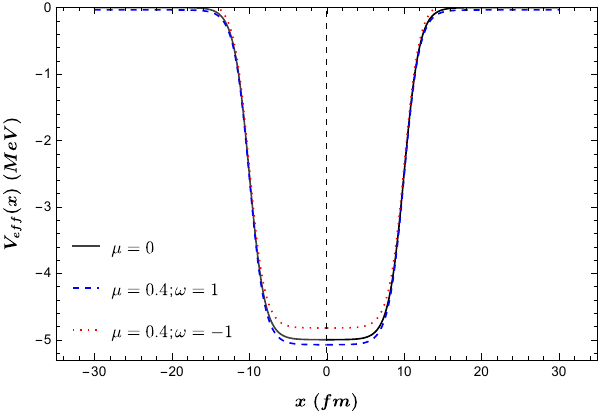}
\caption{Well configuration ($\sigma=-1$).}
\end{subfigure}

\caption{
Approximate effective sWS potentials obtained using the Pekeris-type approximation for representative values of the Dunkl deformation parameter $\mu$ in the two reflection sectors ($\omega=\pm1$). Panels (a) and (b) correspond to the barrier and well configurations, respectively. The calculations were performed using $V_{0}=5~\mathrm{MeV}$, $a=1~\mathrm{fm}^{-1}$, $L=10~\mathrm{fm}$, $mc^{2}=940~\mathrm{MeV}$, and $\hbar c=197.329~\mathrm{MeV\,fm}$.
}
\label{fig:EffectivePotential}
\end{figure}

To illustrate the influence of the Dunkl deformation, the approximate effective potential is shown in Fig.~\ref{fig:EffectivePotential} for several values of the deformation parameter $\mu$ and the reflection eigenvalue $\omega$.

As illustrated in Fig.~\ref{fig:EffectivePotential}, the Dunkl deformation modifies the effective sWS potential in both the barrier and well configurations. The undeformed case ($\mu=0$) is recovered in the absence of the Dunkl deformation, whereas a nonzero value of $\mu$ introduces a sector-dependent inverse-square contribution that distinguishes the effective potentials associated with $\omega=\pm1$. As expected, the effect is most pronounced near the origin, where the inverse-square interaction is strongest. Consequently, both the scattering properties and the bound-state spectrum are expected to exhibit a nontrivial dependence on the Dunkl deformation.

Employing the approximate effective potential given by Eq.~(\ref{eq:ApproximatePotential}), the effective Schr\"odinger equation can be written as
\begin{equation}
\phi''(x) + \frac{2m}{\hbar^2} \Bigg[E - \delta C_0-\frac{\sigma V_0+\delta C_1}{\left(1+e^{a(|x|-L)}\right)} - \frac{\delta C_2}{\left(1+e^{a(|x|-L)}\right)^2}\Bigg] \phi(x) = 0.
\label{eq:EffectiveSchrodinger}
\end{equation}
Equation~(\ref{eq:EffectiveSchrodinger}) forms the basis of the analytical treatment presented in the following subsection.

\subsection{Analytical Solution}

Since the effective potential is symmetric with respect to $x=0$, the solutions of Eq.~(\ref{eq:EffectiveSchrodinger}) have identical functional forms in the regions $x<0$ and $x>0$. It is therefore sufficient to derive the analytical solution for $x<0$, while the corresponding solution for $x>0$ follows directly by symmetry. To transform Eq.~(\ref{eq:EffectiveSchrodinger}) into a hypergeometric-type differential equation, we introduce the dimensionless variable
\begin{equation}
z=\frac{1}{1+e^{-a(x+L)}}. \label{eq:zTransformation}
\end{equation}
The transformation maps the physical domain $(-\infty,0)$ onto the interval
\[
0 \le z \le \frac{1}{1+e^{-aL}} \approx 1,
\]
where the approximation follows from the condition $aL\gg1$.

Substituting Eq.~(\ref{eq:zTransformation}) into Eq.~(\ref{eq:EffectiveSchrodinger}) and introducing the dimensionless parameters
\begin{align}
\epsilon^2 &= -\frac{2m}{a^2\hbar^2}\left(E-\delta C_0\right), \label{eq:EpsilonDefinition}\\
\beta^2 &= -\frac{2m}{a^2\hbar^2}\left(\sigma V_0+\delta C_1\right), \label{eq:BetaDefinition}\\
\gamma^2 &= -\frac{2m}{a^2\hbar^2}\delta C_2.
\label{eq:GammaDefinition}
\end{align}
one obtains
\begin{equation}
\phi''(z)+ \frac{1-2z}{z(1-z)}\phi'(z)+\frac{
-\epsilon^2+\beta^2 z+\gamma^2 z^2}{z^2(1-z)^2}\phi(z)=0.\label{eq:zEquation}
\end{equation}
The transformed differential equation has regular singular points at $z=0$ and $z=1$. Hence, we seek the solution in the form
\begin{equation}
\phi(z) = z^{\epsilon} (1-z)^{\nu} f(z),
\label{eq:Ansatz}
\end{equation}
where
\begin{equation}
\nu = \sqrt{\epsilon^2-\beta^2-\gamma^2}=\sqrt{-\frac{2m}{a^2 \hbar^2}\big[E-\sigma V_0- \delta \left(C_0+C_1+C_2\right)\big]}\, . \label{eq:NuDefinition}
\end{equation}
Substituting Eq.~(\ref{eq:Ansatz}) into Eq.~(\ref{eq:zEquation}) yields
\begin{equation}
z(1-z)f''(z) +\Big[c_h-(1+a_h+b_h)z\Big]f'(z)-a_h b_h\,f(z)=0, \label{eq:HyperEquation}
\end{equation}
which is the standard Gauss hypergeometric differential equation. Its general solution is therefore given by
\begin{equation}
f(z)= A\,{}_2F_1(a_h,b_h,c_h;z)+B\,z^{-2\epsilon}{}_2F_1 \left(1+a_h-c_h,\, 1+b_h-c_h,\, 2-c_h;\,z \right), \label{eq:HyperSolution}
\end{equation}
where the hypergeometric parameters are
\begin{equation}
a_h=\epsilon+\theta+\nu, \qquad
b_h=1+\epsilon-\theta+\nu,
\qquad
c_h=1+2\epsilon,
\label{eq:HyperParameters}
\end{equation}
with
\begin{equation}
\theta = \frac{1}{2} \pm \sqrt{ \frac{1}{4} -\gamma^{2} }. \label{eq:ThetaDefinition}
\end{equation}
For the parameter range considered in this work, the quantity $\theta$ remains real. The parameters $a_h$, $b_h$, and $c_h$ completely determine the hypergeometric solution. In the following sections, the general solution is specialized to the bound-state and scattering-state problems by imposing the appropriate boundary conditions. Throughout the subsequent analysis, the subscripts $b$ and $s$ are used to denote the corresponding bound-state and scattering-state quantities, respectively.

\section{Bound States}\label{Sec:sec3}

We first examine the bound-state solutions corresponding to the attractive sWS well configuration ($\sigma=-1$). Since bound states satisfy $E_b<0$ (and hence $E_b<\delta C_0$), it follows that $\epsilon_b^2>0$. Accordingly, the positive branch is selected as
\begin{equation}
\epsilon_b\equiv k_b=
\sqrt{-\frac{2m}{a^2\hbar^2}
\left(E_b-\delta C_0\right)}.
\label{eq:BoundEpsilon}
\end{equation}
Furthermore, the parameter $\nu_b$ is introduced through
\begin{equation}
\nu_b \equiv i\kappa_b,
\label{eq:BoundNu}
\end{equation}
where
\begin{equation}
\kappa_b=
\sqrt{
\frac{2m}{a^2\hbar^2}
\left[
E_b+V_0-\delta(C_0+C_1+C_2)
\right]
}.
\label{eq:BoundKappa}
\end{equation}
Substituting Eqs.~\eqref{eq:BoundEpsilon} and \eqref{eq:BoundNu} into the general hypergeometric solution derived in Section~2 yields
\begin{align}
\phi_b(z)
&=
A_b
z^{k_b}
(1-z)^{i\kappa_b}
{}_2F_1(a_{hb},b_{hb},c_{hb};z)
\nonumber\\
&\quad+
B_b
z^{-k_b}
(1-z)^{i\kappa_b}
{}_2F_1
\left(
1+a_{hb}-c_{hb},
1+b_{hb}-c_{hb},
2-c_{hb};z
\right).
\label{eq:BoundGeneralSolution}
\end{align}
In the $x<0$ region, imposing the asymptotic condition requires $B_{b,l}=0$. Therefore, the physically acceptable solution in the left region is given by
\begin{equation}
\phi_{b,l}(x)=A_{b,l}
\left(\frac{1}{1+e^{-a(x+L)}}\right)^{k_b}
\left(\frac{1}{1+e^{a(x+L)}}\right)^{i\kappa_b}
{}_2F_1\bigg(a_{hb},b_{hb},c_{hb};
\frac{1}{1+e^{-a(x+L)}}\bigg).
\label{eq:PhysicalLeftBoundSolution}
\end{equation}
Similarly, in the $x>0$ region, the asymptotic boundary condition removes the exponentially divergent contribution, leading to
\begin{equation}
\phi_{b,r}(x)=A_{b,r}
\left(\frac{1}{1+e^{a(x-L)}}\right)^{k_b}
\left(\frac{1}{1+e^{-a(x-L)}}\right)^{i\kappa_b}
{}_2F_1\bigg(a_{hb},b_{hb},c_{hb};
\frac{1}{1+e^{a(x-L)}}\bigg).
\label{eq:PhysicalRightBoundSolution}
\end{equation}
The bound-state wave functions obtained in the left and right regions must satisfy the continuity conditions at the symmetry point $x=0$:
\begin{equation}
\phi_{b,l}(0)=\phi_{b,r}(0),
\qquad
\phi_{b,l}'(0)=\phi_{b,r}'(0).
\label{eq:BoundContinuity}
\end{equation}
Substituting the left- and right-region solutions,
Eqs.~\eqref{eq:PhysicalLeftBoundSolution} and
\eqref{eq:PhysicalRightBoundSolution},
into the continuity conditions yields
\begin{equation}
\left(A_{b,l}-A_{b,r}\right)
\left(\frac{1}{1+e^{-aL}}\right)^{k_b}
\Lambda^{\nu_b}
\left(
N_1\Lambda^{-2\nu_b}+N_2
\right)=0,
\label{eq:BoundWaveFuncsEqualsAt0}
\end{equation}
and
\begin{equation}
a\left(A_{b,l}+A_{b,r}\right)
\left(\frac{1}{1+e^{-aL}}\right)^{k_b}
\Lambda^{\nu_b-1}
\left(
M_1\Lambda^{-2\nu_b}+M_2
\right)=0,
\label{eq:BoundWaveDerivativeEqualsAt0}
\end{equation}
with
\begin{equation}
\Lambda=1+e^{aL}.
\label{eq:Lambda}
\end{equation}
Here, the coefficients $N_1$, $N_2$, $M_1$, and $M_2$ are defined by

%%%%%%%%
\begin{equation}
\begin{aligned}
N_1&=
\frac{\Gamma(2k_b+1)\Gamma(-2i\kappa_b)}
{\Gamma(k_b-\theta-i\kappa_b+1)
 \Gamma(k_b+\theta-i\kappa_b)}, &
N_2&=
\frac{\Gamma(2k_b+1)\Gamma(2i\kappa_b)}
{\Gamma(k_b+\theta+i\kappa_b)
 \Gamma(k_b-\theta+i\kappa_b+1)},
\\[2mm]
M_1&=
N_1\left(k_b-i\kappa_be^{aL}\right),
&
M_2&=
N_2\left(k_b+i\kappa_be^{aL}\right).
\end{aligned}
\label{eq:N12M12Coefficients}
\end{equation}

Since the effective potential is symmetric under the spatial transformation $x\rightarrow -x$, the bound-state solutions are classified according to their spatial parity. This classification is independent of the Dunkl reflection sector, so that each sector may accommodate both even- and odd-parity solutions, depending on the imposed boundary conditions. The continuity conditions, Eqs.~\eqref{eq:BoundWaveFuncsEqualsAt0} and \eqref{eq:BoundWaveDerivativeEqualsAt0}, are then imposed under the odd- and even-parity constraints, $A_{b,l}=-A_{b,r}$ and $A_{b,l}=A_{b,r}$, respectively.

\begin{equation}
E_{b,n}^{\mathrm{odd}} = \delta C_0 - \frac{a^2\hbar^2}{2m} \left[ \beta_b^{\,2} + \gamma^{\,2}+ \left(\frac{\ln N_1-\ln N_2+i\pi(2n+1)}{2\ln\Lambda}\right)^2\right]. \label{eq:OddSolutionForEigenValue}
\end{equation}
\begin{equation}
E_{b,n}^{\mathrm{even}} = \delta C_0 -\frac{a^2\hbar^2}{2m}\left[\beta_b^{\,2}+\gamma^{\,2}+\left(\frac{\ln M_1-\ln M_2+i\pi(2n+1)}{2\ln\Lambda}\right)^2\right]. \label{eq:EvenSolutionForEigenValue}
\end{equation}
Throughout the above expressions, the logarithm is evaluated on a consistent branch. Consequently, the quantities $\ln N_1-\ln N_2$ and $\ln M_1-\ln M_2$ should be understood as branch-consistent logarithmic differences rather than the principal-branch expressions $\ln(N_1/N_2)$ and $\ln(M_1/M_2)$.

The corresponding bound-state eigenfunctions are obtained by substituting $E_{b,n}^{\mathrm{odd}}$ and $E_{b,n}^{\mathrm{even}}$ into Eqs.~\eqref{eq:PhysicalLeftBoundSolution} and \eqref{eq:PhysicalRightBoundSolution}. Ehe odd-parity eigenfunctions are given by
\small
\begin{equation}
\phi_{b,l;n}^{\mathrm{odd}}(x)=A_{b,l} \left(\frac{1}{1+e^{-a(x+L)}}\right)^{k_{b,n}^{\mathrm{odd}}} \left(\frac{1}{1+e^{a(x+L)}}\right)^{i\kappa_{b,n}^{\mathrm{odd}}} {}_2F_1\!\left(a_{hb},b_{hb},c_{hb};\frac{1}{1+e^{-a(x+L)}}\right),\label{eq:OddBoundLeftEigenfunction}
\end{equation} \normalsize
and \small
\begin{equation}
\phi_{b,r;n}^{\mathrm{odd}}(x)=-A_{b,l} \left(\frac{1}{1+e^{a(x-L)}}\right)^{k_{b,n}^{\mathrm{odd}}}\left(\frac{1}{1+e^{-a(x-L)}}\right)^{i\kappa_{b,n}^{\mathrm{odd}}}{}_2F_1\!\left(a_{hb},b_{hb},c_{hb};\frac{1}{1+e^{a(x-L)}}\right), \label{eq:OddBoundRightEigenfunction}
\end{equation} \normalsize

Similarly, the left- and right-region eigenfunctions for the even-parity states are
\small
\begin{equation}
\phi_{b,l;n}^{\mathrm{even}}(x)=A_{b,l} \left(\frac{1}{1+e^{-a(x+L)}}\right)^{k_{b,n}^{\mathrm{even}}} \left(\frac{1}{1+e^{a(x+L)}}\right)^{i\kappa_{b,n}^{\mathrm{even}}} {}_2F_1\!\left(a_{hb},b_{hb},c_{hb};\frac{1}{1+e^{-a(x+L)}}\right),\label{eq:EvenBoundLeftEigenfunction}
\end{equation} \normalsize
and \small
\begin{equation}
\phi_{b,r;n}^{\mathrm{even}}(x)=A_{b,l}\left(\frac{1}{1+e^{a(x-L)}}\right)^{k_{b,n}^{\mathrm{even}}}\left(\frac{1}{1+e^{-a(x-L)}}\right)^{i\kappa_{b,n}^{\mathrm{even}}}{}_2F_1\!\left(a_{hb},b_{hb},c_{hb};\frac{1}{1+e^{a(x-L)}}\right),\label{eq:EvenBoundRightEigenfunction}
\end{equation} \normalsize

\subsection{Illustrative Results}

To examine the influence of the Dunkl parameter on the bound-state spectrum, the odd- and even-parity energy eigenvalues were computed numerically for different values of $\mu$. Unless otherwise stated, all numerical calculations were performed using the parameter set $V_0=100~\mathrm{MeV}$, $a=1~\mathrm{fm}^{-1}$, $L=10~\mathrm{fm}$, $mc^2=940~\mathrm{MeV}$, and $\hbar c=197.329~\mathrm{MeV\,fm}$. The calculated bound-state energies are listed in Tables~\ref{tab:BoundOddEnergies} and \ref{tab:BoundEvenEnergies}, corresponding to the odd- and even-parity solutions, respectively. For each quantum number $n$, the ordered pair $(n^{+},n^{-})$ denotes the energy eigenvalues associated with the reflection eigenvalues $\omega=+1$ and $\omega=-1$, respectively.

\clearpage

\begin{landscape}

\thispagestyle{plain}
\begin{table}[t]
\centering
\caption{Bound-state energy eigenvalues (MeV) for the odd-parity solutions. For each quantum number $n$, the ordered pair $(n^{+},n^{-})$ corresponds to the reflection eigenvalues $\omega=+1$ and $\omega=-1$, respectively. }

\label{tab:BoundOddEnergies}

\renewcommand{\arraystretch}{1.15}
\setlength{\tabcolsep}{6pt}

\begin{tabular}{cccccccc}
\toprule
$\mu$
&
$(0^{+},0^{-})$
&
$(1^{+},1^{-})$
&
$(2^{+},2^{-})$
&
$(3^{+},3^{-})$
&
$(4^{+},4^{-})$
&
$(5^{+},5^{-})$
&
$(6^{+},6^{-})$
\\
\midrule

0.0 &
(-96.581,-96.581) &
(-88.219,-88.219) &
(-76.492,-76.492) &
(-62.235,-62.235) &
(-46.196,-46.196) &
(-29.285,-29.285) &
(-12.958,-12.958)
\\

0.1 &
(-96.609,-96.546) &
(-88.246,-88.185) &
(-76.519,-76.459) &
(-62.261,-62.203) &
(-46.221,-46.165) &
(-29.309,-29.256) &
(-12.980,-12.931)
\\

0.2 &
(-96.631,-96.506) &
(-88.268,-88.145) &
(-76.540,-76.420) &
(-62.282,-62.165) &
(-46.241,-46.128) &
(-29.328,-29.221) &
(-12.997,-12.899)
\\

0.3 &
(-96.646,-96.459) &
(-88.283,-88.099) &
(-76.555,-76.374) &
(-62.296,-62.121) &
(-46.255,-46.086) &
(-29.342,-29.181) &
(-13.009,-12.862)
\\

0.4 &
(-96.656,-96.405) &
(-88.293,-88.046) &
(-76.564,-76.323) &
(-62.305,-62.071) &
(-46.264,-46.038) &
(-29.350,-29.135) &
(-13.017,-12.820)
\\

\bottomrule
\end{tabular}
\end{table}

\vspace{0.7cm}

\begin{table}[b]
\centering
\caption{Bound-state energy eigenvalues (MeV) for the even-parity solutions. For each quantum number $n$, the ordered pair $(n^{+},n^{-})$ corresponds to the reflection eigenvalues $\omega=+1$ and $\omega=-1$, respectively.}
\label{tab:BoundEvenEnergies}

\renewcommand{\arraystretch}{1.15}
\setlength{\tabcolsep}{6pt}

\begin{tabular}{cccccccc}
\toprule
$\mu$
&
$(0^{+},0^{-})$
&
$(1^{+},1^{-})$
&
$(2^{+},2^{-})$
&
$(3^{+},3^{-})$
&
$(4^{+},4^{-})$
&
$(5^{+},5^{-})$
&
$(6^{+},6^{-})$
\\
\midrule

0.0 &
(-99.077,-99.077) &
(-92.886,-92.886) &
(-82.718,-82.718) &
(-69.634,-69.634) &
(-54.388,-54.388) &
(-37.779,-37.779) &
(-20.914,-20.914)
\\

0.1 &
(-99.105,-99.042) &
(-92.914,-92.852) &
(-82.745,-82.684) &
(-69.661,-69.601) &
(-54.414,-54.356) &
(-37.803,-37.748) &
(-20.937,-20.885)
\\

0.2 &
(-99.127,-99.001) &
(-92.936,-92.811) &
(-82.767,-82.645) &
(-69.681,-69.562) &
(-54.434,-54.318) &
(-37.823,-37.712) &
(-20.955,-20.852)
\\

0.3 &
(-99.143,-98.954) &
(-92.951,-92.765) &
(-82.782,-82.599) &
(-69.696,-69.518) &
(-54.448,-54.275) &
(-37.837,-37.671) &
(-20.968,-20.813)
\\

0.4 &
(-99.152,-98.901) &
(-92.960,-92.712) &
(-82.791,-82.547) &
(-69.705,-69.467) &
(-54.457,-54.226) &
(-37.845,-37.624) &
(-20.976,-20.769)
\\

\bottomrule
\end{tabular}
\end{table}

\end{landscape}

\clearpage

\begin{figure}[htb!]
\centering

\begin{subfigure}{0.48\textwidth}
\centering
\includegraphics[width=\linewidth]{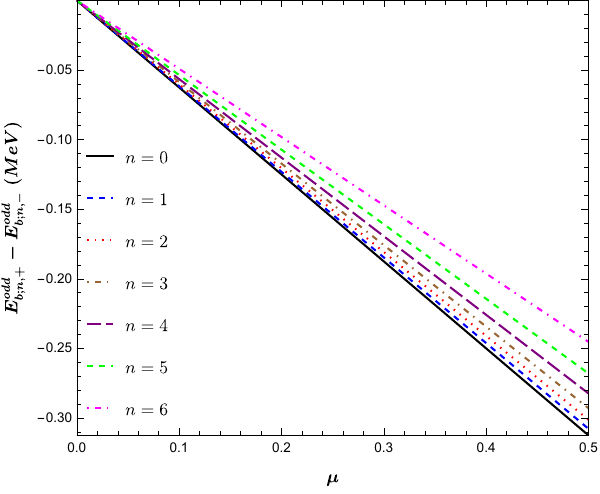}
\caption{Odd-parity states.}
\end{subfigure}
\hfill
\begin{subfigure}{0.48\textwidth}
\centering
\includegraphics[width=\linewidth]{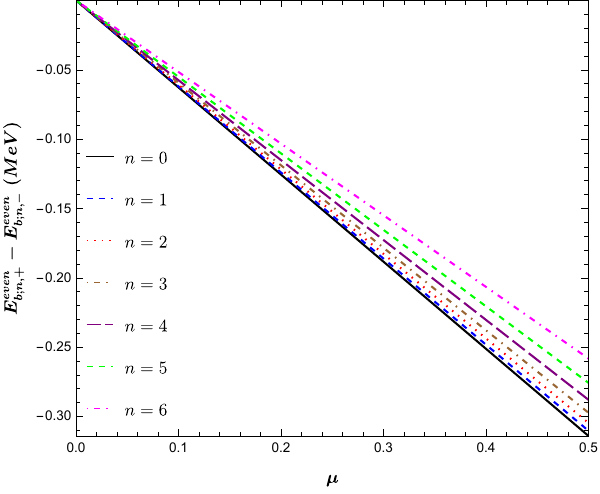}
\caption{Even-parity states.}
\end{subfigure}

\caption{
Variation of the bound-state energy difference
$\Delta E_{b,n}=E_{b,n,+}-E_{b,n,-}$ as a function of the Dunkl deformation parameter $\mu$ for the first three bound states. Panels (a) and (b) correspond to the odd- and even-parity solutions, respectively.} 
\label{fig:BoundEnDifS}
\end{figure}

\newpage
To further examine the influence of the Dunkl parameter on the bound-state spectrum, Fig.~\ref{fig:BoundEnDifS} shows the variation of the energy difference between the reflection sectors, $E_{b,n,+}-E_{b,n,-}$, as a function of $\mu$ for the odd- and even-parity bound states.

As shown in Fig.~\ref{fig:BoundEnDifS}, the energy difference vanishes at $\mu=0$, corresponding to the ordinary quantum-mechanical limit. For both odd- and even-parity states, the energy separation increases monotonically with increasing $\mu$. Although the magnitude of the energy difference depends weakly on the bound-state quantum number, all bound states display the same qualitative behavior, indicating that the Dunkl parameter affects both parity sectors in a similar manner.

\begin{figure}[htb!]
\centering

% --- upper row ---
\begin{subfigure}{0.48\textwidth}
\centering
\includegraphics[width=\linewidth]{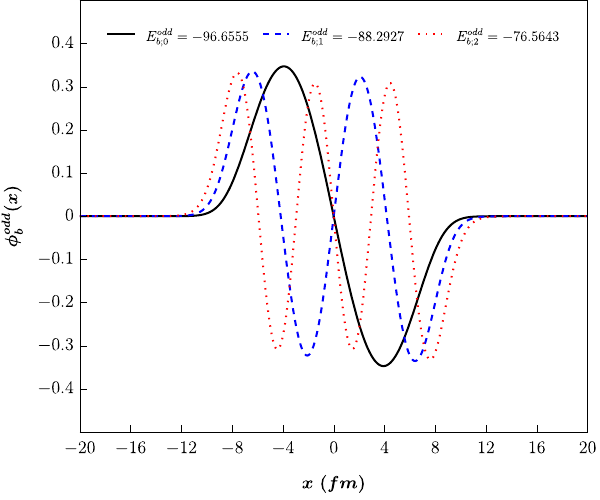}
\caption{Normalized odd-parity bound-state eigenfunctions.}
\end{subfigure}
\hfill
\begin{subfigure}{0.48\textwidth}
\centering
\includegraphics[width=\linewidth]{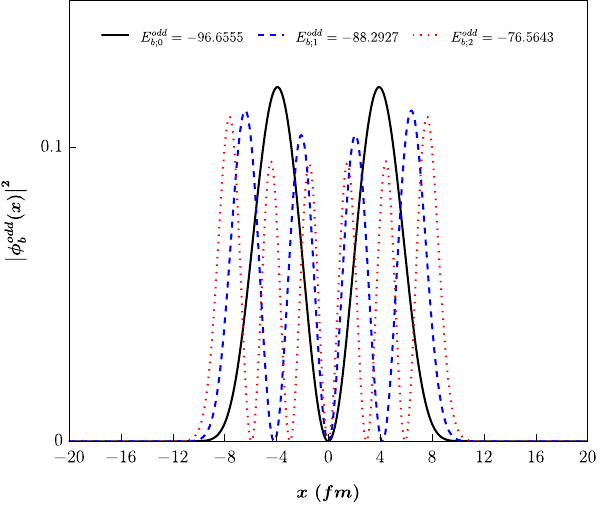}
\caption{Probability densities for the odd-parity bound states.}
\end{subfigure}

\vspace{0.8cm}

% --- lower row ---
\begin{subfigure}{0.48\textwidth}
\centering
\includegraphics[width=\linewidth]{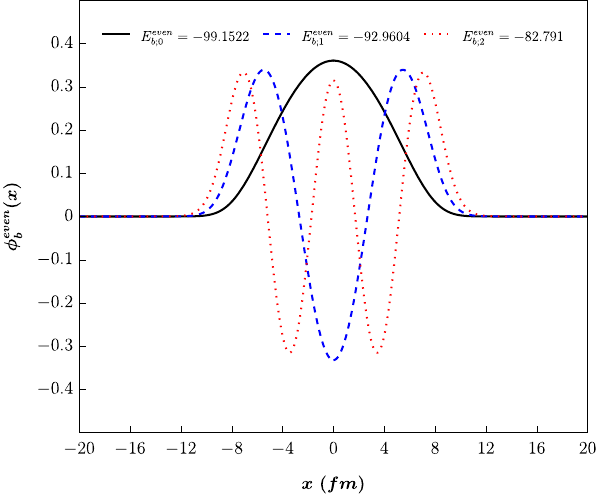}
\caption{Normalized even-parity bound-state eigenfunctions.}
\end{subfigure}
\hfill
\begin{subfigure}{0.48\textwidth}
\centering
\includegraphics[width=\linewidth]{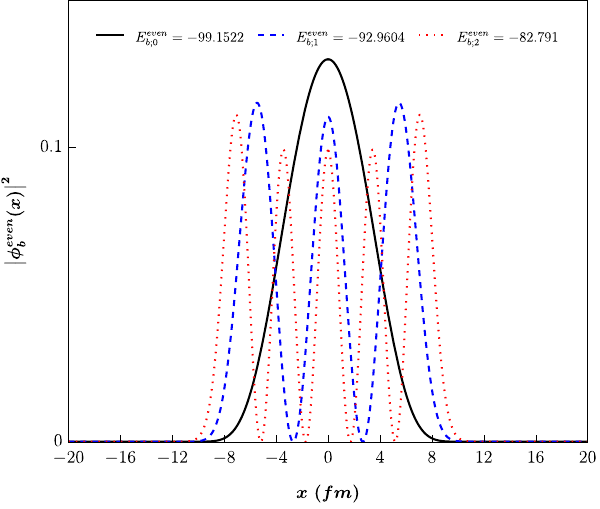}
\caption{Probability densities for the even-parity bound states.}
\end{subfigure}

\caption{
Normalized bound-state eigenfunctions and the corresponding probability densities for the effective sWS potential for $\mu=0.4$ and $\omega=1$. Panels (a) and (b) correspond to the odd-parity eigenfunctions and their probability densities, respectively, whereas panels (c) and (d) show the corresponding even-parity results.} \label{fig:BoundWaveAndProbPlots}
\end{figure}

Having examined the evolution of the bound-state energy spectrum, we next consider the spatial properties of the corresponding eigenfunctions and their probability densities, which are displayed in Fig.~\ref{fig:BoundWaveAndProbPlots}.

\newpage 
As expected, the odd-parity wave functions shown in Fig.~\ref{fig:BoundWaveAndProbPlots}(a) vanish at the symmetry point $x=0$, whereas the even-parity wave functions in Fig.~\ref{fig:BoundWaveAndProbPlots}(c) remain finite at the center of the potential. In both parity sectors, the number of nodes increases with the bound-state quantum number, reflecting the increasing excitation level.

The corresponding probability densities, shown in Figs.~\ref{fig:BoundWaveAndProbPlots}(b) and \ref{fig:BoundWaveAndProbPlots}(d), are symmetric with respect to $x=0$, consistent with the spatial symmetry of the effective potential. Furthermore, the probability densities are localized within the potentialwell and decay rapidly outside the interaction region, confirming the
bound-state nature of the obtained solutions.

\begin{figure}[htb!]
\centering

\begin{subfigure}{0.48\textwidth}
\centering
\includegraphics[width=\linewidth]{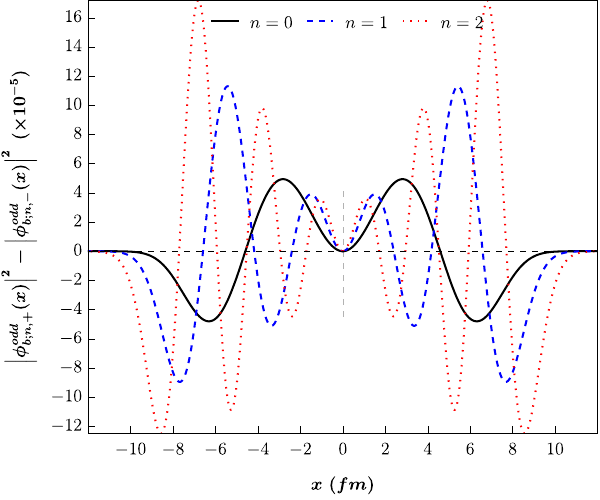}
\caption{Difference between the probability densities of the odd-parity bound states for the reflection eigenvalues $\omega=+1$ and $\omega=-1$. Here, $\mu=0.4$.}
\end{subfigure}
\hfill
\begin{subfigure}{0.48\textwidth}
\centering
\includegraphics[width=\linewidth]{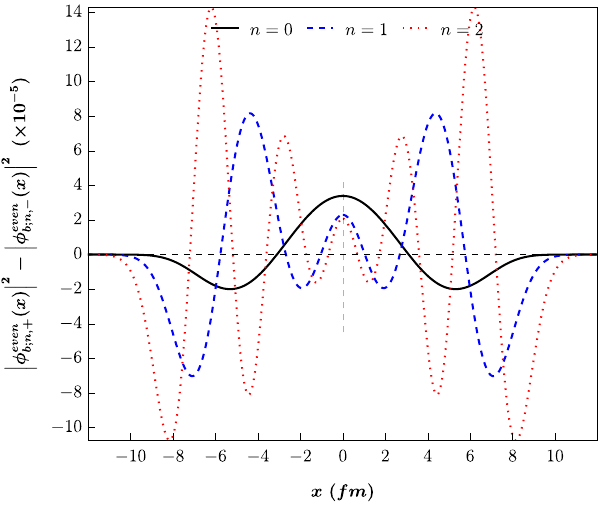}
\caption{Difference between the probability densities of the even-parity bound states for the reflection eigenvalues $\omega=+1$ and $\omega=-1$.}
\end{subfigure}

\caption{
Difference between the probability densities corresponding to the reflection
eigenvalues $\omega=+1$ and $\omega=-1$ for the first five bound states.
Panels (a) and (b) correspond to the odd- and even-parity solutions,
respectively. }
\label{fig:BoundProbDifS}
\end{figure}

Although the bound-state eigenfunctions exhibit the expected spatial parity properties, the influence of the Dunkl reflection degree of freedom is more clearly manifested in the corresponding probability densities. To this end, Fig.~\ref{fig:BoundProbDifS} presents the spatial distribution of the probability-density difference between the $\omega=+1$ and $\omega=-1$ reflection sectors for the first three odd- and even-parity bound states.

Since the plotted quantity represents the difference between two probability densities, both positive and negative regions appear, indicating the spatial intervals where the $\omega=+1$ and $\omega=-1$ reflection sectors possess different localization probabilities. In both parity sectors, the largest deviations are observed within the effective potential well, whereas the difference gradually vanishes outside the interaction region owing to the exponential decay of the bound-state wave functions. Furthermore, the spatial profile of the probability-density difference reflects the nodal structure of the corresponding bound-state eigenfunctions. These results demonstrate that the reflection eigenvalue associated with the Dunkl operator influences the spatial probability distribution of the bound states without altering their odd- or even-parity character.

\section{Scattering States}\label{Sec:sec4}

We next investigate the scattering-state solutions corresponding to the repulsive effective sWS barrier configuration ($\sigma=+1$). Unlike the bound-state case, the scattering solutions are characterized by oscillatory asymptotic behavior. Accordingly, the parameter $\epsilon_s$ is introduced as
\begin{equation}
\epsilon_s\equiv ik_s, \label{eq:ScatteringEpsilon}
\end{equation}
where
\begin{equation}
k_s= \sqrt{ \frac{2m}{a^2\hbar^2}\left(E_s-\delta C_0\right)}. \label{eq:ScatteringK}
\end{equation}
The requirement that the asymptotic wave number be real restricts the scattering energy to $E_s>\delta C_0$. Furthermore, the parameter $\nu_s$ is introduced through
\begin{equation}
\nu_s\equiv i\kappa_s, \label{eq:ScatteringNu}
\end{equation}
where
\begin{equation}
\kappa_s= \sqrt{\frac{2m}{a^2\hbar^2} \Big[ E_s-V_0-\delta\left(C_0+C_1+C_2\right)\Big]}. \label{eq:ScatteringKappa}
\end{equation}
With the above definitions, the general hypergeometric solution assumes the form
\begin{align}
\phi_s(z) &= A_s z^{ik_s} (1-z)^{i\kappa_s} {}_2F_1(a_{hs},b_{hs},c_{hs};z) \nonumber\\ 
&\quad + B_s z^{-ik_s} (1-z)^{i\kappa_s} {}_2F_1 \left( 1+a_{hs}-c_{hs}, 1+b_{hs}-c_{hs}, 2-c_{hs};z \right). \label{eq:ScatteringGeneralSolution}
\end{align}
Assuming that the incident wave propagates from the left, the asymptotic boundary conditions require that the left region contains both the incident and reflected waves. For this reason the asymptotic form of the left-region wave function is given by
\begin{equation}
\phi_{s,l}(x\rightarrow-\infty)=A_{s,l}e^{ik_sa(x+L)} + B_{s,l}e^{-ik_sa(x+L)}, \label{eq:ScatteringWaveFuncLeftSideMinusInfty}
\end{equation}
Similarly, the absence of an incident wave from the right implies that only the transmitted wave exists in the right asymptotic region. Therefore, by setting $A_{s,r}=0$, the asymptotic form of the right-region wave function is obtained as
\begin{equation}
\phi_{s,r}(x\rightarrow+\infty) = B_{s,r}e^{ik_sa(x-L)}. \label{eq:ScatteringWaveFuncRightSidePlusInfty}
\end{equation}
%%%%%%%%%%%%%%%
Evaluating the left- and right-region wave functions in the limit $x\rightarrow0$ gives
\begin{equation}
\begin{aligned}
\phi_{s,l}(x\rightarrow0)
&= \left(A_{s,l}N_1+B_{s,l}N_3\right)e^{-\nu_sa(x+L)}
+ \left(A_{s,l}N_2+B_{s,l}N_4\right)e^{\nu_sa(x+L)},
\\[2mm]
\phi_{s,r}(x\rightarrow0)
&= B_{s,r} \left( N_3e^{\nu_sa(x-L)} + N_4e^{-\nu_sa(x-L)} \right),
\end{aligned} \label{eq:ScatteringWaveFunctionsZero}
\end{equation}
with the corresponding first derivatives
\begin{equation}
\begin{aligned}
\phi'_{s,l}(x\rightarrow0)
&= (\nu_sa) \left[ -\left(A_{s,l}N_1+B_{s,l}N_3\right)e^{-\nu_sa(x+L)}
+ \left(A_{s,l}N_2+B_{s,l}N_4\right)e^{\nu_sa(x+L)} \right], \\[2mm]
\phi'_{s,r}(x\rightarrow0)
&= (\nu_sa)B_{s,r} \left( N_3e^{\nu_sa(x-L)}- N_4e^{-\nu_sa(x-L)}\right),
\end{aligned} \label{eq:ScatteringWaveFunctionDerivativesZero}
\end{equation}
where the coefficients $N_3$ and $N_4$ are defined by
\begin{equation}
\begin{aligned}
N_3&= \frac{\Gamma(1-2ik_s)\Gamma(-2i\kappa_s)}{\Gamma(-ik_s-\theta-i\kappa_s+1) \Gamma(-ik_s+\theta-i\kappa_s)}, \\[2mm]
N_4&=\frac{\Gamma(1-2ik_s)\Gamma(2i\kappa_s)}{\Gamma(-ik_s+\theta+i\kappa_s) \Gamma(-ik_s-\theta+i\kappa_s+1)}.
\end{aligned} \label{eq:N34Coefficients}
\end{equation}
The coefficients $N_1$ and $N_2$ retain the forms given in Eq.~\eqref{eq:N12M12Coefficients}, with the replacements $k_b\rightarrow ik_s$ and $\kappa_b\rightarrow\kappa_s$.

\subsection{Reflection and Transmission Coefficients}

The reflection and transmission coefficients are determined from the
probability current density,
\begin{equation}
j(x)=
\frac{i\hbar}{2m}
\left[
\phi_s(x)\frac{d\phi_s^{*}(x)}{dx}
-
\phi_s^{*}(x)\frac{d\phi_s(x)}{dx}
\right].
\label{eq:ProbabilityCurrent}
\end{equation}
Using the asymptotic wave functions,
Eqs.~\eqref{eq:ScatteringWaveFuncLeftSideMinusInfty} and
\eqref{eq:ScatteringWaveFuncRightSidePlusInfty},
the reflection and transmission coefficients are defined by
\begin{equation}
R=
\left|
\frac{j_{\mathrm{ref}}}{j_{\mathrm{inc}}}
\right|
=
\left|
\frac{B_{s,l}}{A_{s,l}}
\right|^2,
\label{eq:ReflectionCoefficient1}
\end{equation}

\begin{equation}
T=
\left|
\frac{j_{\mathrm{trans}}}{j_{\mathrm{inc}}}
\right|
=
\left|
\frac{B_{s,r}}{A_{s,l}}
\right|^2.
\label{eq:TransmissionCoefficient1}
\end{equation}

To obtain explicit expressions for the reflection and transmission coefficients, the continuity conditions at the symmetry point $x=0$ are imposed, yielding
\begin{equation}
R=
\left|
\frac{B_{s,l}}{A_{s,l}}
\right|^2
=
\left|
\frac{N_2N_4e^{\,4i a  \kappa_s L}-N_1N_3}
{N_3^{\,2}-N_4^{\,2}e^{\,4i a  \kappa_s L}}
\right|^2,
\label{eq:ReflectionCoefficient2}
\end{equation}

\begin{equation}
T=
\left|
\frac{B_{s,r}}{A_{s,l}}
\right|^2
=
\left|
\frac{\left(N_2N_3-N_1N_4\right)e^{\,2 i a  \kappa_s L}}
{N_3^{\,2}-N_4^{\,2}e^{\,4i a  \kappa_s L}}
\right|^2.
\label{eq:TransmissionCoefficient2}
\end{equation}
For $E_s\leq V_0$, the parameter $\nu_s$ is real, leading to the relations $N_1=N_3^{*}$ and $N_2=N_4^{*}$. In contrast, for $E_s>V_0$, $\nu_s$ becomes purely imaginary, yielding $N_1=N_4^{*}$ and $N_2=N_3^{*}$. In both energy regimes, the reflection and transmission coefficients satisfy
\begin{equation}
R+T=1,
\end{equation}
confirming the conservation of the probability current.

\subsection{Illustrative Results}

\begin{figure}[htb!]
	\centering
	
	\begin{subfigure}{0.48\textwidth}
		\centering
		\includegraphics[width=\textwidth]{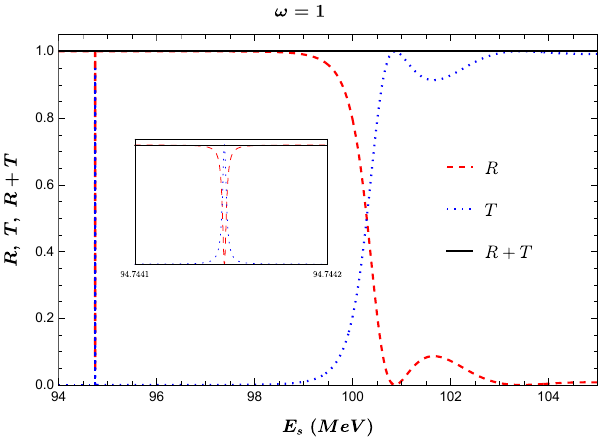}
		\caption*{(a)}
	\end{subfigure}
	\hfill
	\begin{subfigure}{0.48\textwidth}
		\centering
		\includegraphics[width=\textwidth]{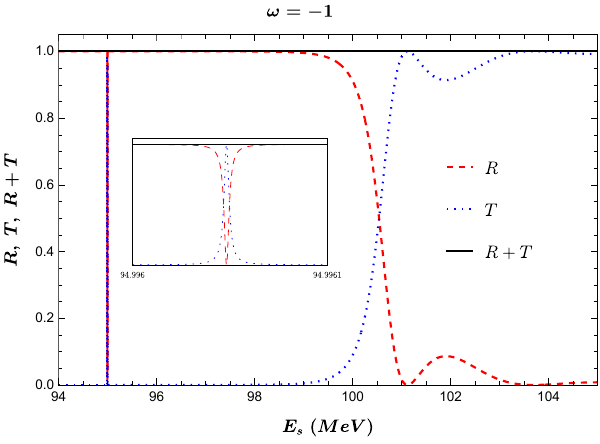}
		\caption*{(b)}
	\end{subfigure}
	
	\vspace{0.8cm}
	
	\begin{subfigure}{0.6\textwidth}
		\centering
		\includegraphics[width=\textwidth]{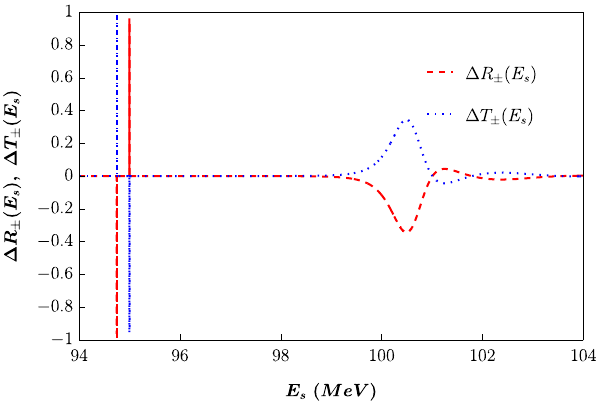}
		\caption*{(c)}
	\end{subfigure}
	
	\caption{
Reflection and transmission coefficients as functions of the scattering energy for the effective Dunkl--Woods--Saxon barrier for $\mu=0.4$.
Panels (a) and (b) display the reflection and transmission coefficients for the reflection sectors $\omega=+1$ and $\omega=-1$, respectively.
The vertical dashed line indicates the barrier height $V_0$.
Panel (c) presents the differences
$\Delta R(E_s)=R_{+}(E_s)-R_{-}(E_s)$
and
$\Delta T(E_s)=T_{+}(E_s)-T_{-}(E_s)$,
demonstrating the influence of the reflection sector on the scattering properties.
}
	\label{fig:ScatteringPlots}
\end{figure}

%The analytical expressions derived in the previous subsection are now employed to investigate the scattering properties of the effective sWS barrier. For consistency, the same parameter values as those adopted in the bound-state analysis are used throughout this subsection. The reflection and transmission coefficients obtained from Eqs.~\eqref{eq:ReflectionCoefficient2} and \eqref{eq:TransmissionCoefficient2} are presented in Fig.~\ref{fig:ScatteringPlots}.
The analytical expressions for the reflection and transmission coefficients obtained in Eqs.~\eqref{eq:ReflectionCoefficient2} and \eqref{eq:TransmissionCoefficient2} are now employed to investigate the scattering properties of the effective sWS barrier. To facilitate a direct comparison with the bound-state results presented in Sec.~3.1, the same set of model parameters is adopted. The corresponding reflection and transmission coefficients are shown in Fig.~\ref{fig:ScatteringPlots}.

\newpage Figure~\ref{fig:ScatteringPlots} illustrates the energy dependence of the reflection and transmission coefficients for the effective sWS barrier in the two reflection sectors, $\omega=\pm1$. As the incident energy increases, the reflection probability decreases continuously, whereas the transmission probability exhibits the complementary behavior, approaching unity at sufficiently high energies. This evolution reflects the gradual reduction of the effective barrier opacity as the kinetic energy of the incident particle increases.

Although the overall scattering characteristics remain qualitatively similar in both reflection sectors, the Dunkl deformation introduces noticeable quantitative differences through the reflection-dependent inverse-square contribution to the effective potential. These differences are more clearly illustrated in Fig.~\ref{fig:ScatteringPlots}(c), where the quantities $\Delta R(E)=R_{+}(E)-R_{-}(E)$ and $\Delta T(E)=T_{+}(E)-T_{-}(E)$ exhibit their largest deviations within a relatively narrow energy interval. Outside this region, the two reflection sectors gradually converge, indicating that the influence of the Dunkl deformation becomes less significant away from the transition region.

A remarkable feature of the transmission spectrum is the appearance of a narrow resonance-like structure below the barrier height. The existence of such a localized enhancement suggests the formation of a temporary trapping state inside the effective barrier. Since the scattering coefficients alone do not provide a complete characterization of this phenomenon, a detailed analysis based on complex-energy solutions is presented in the following subsection.

\begin{figure}[htp!]
	\centering
	
	% --- üst satır ---
	\begin{subfigure}{0.48\textwidth}
		\centering
		\includegraphics[width=\textwidth]{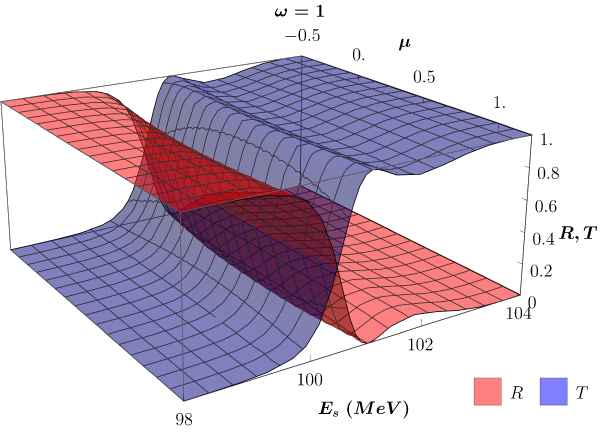}
		\caption*{(a)}
	\end{subfigure}
	\hfill
	\begin{subfigure}{0.48\textwidth}
		\centering
		\includegraphics[width=\textwidth]{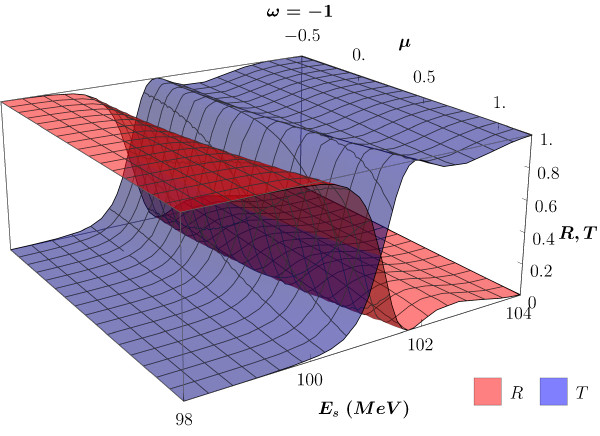}
		\caption*{(b)}
	\end{subfigure}
	
	\caption{
Three-dimensional representations of the reflection coefficient $R$ and the transmission coefficient $T$ as functions of the Dunkl deformation parameter $\mu$ and the scattering energy $E_s$. Panels (a) and (b) correspond to the reflection sectors $\omega=+1$ and $\omega=-1$, respectively. 
}
	\label{fig:Scattering3D}
\end{figure}

Figure~\ref{fig:Scattering3D} provides a three-dimensional visualization of the reflection and transmission coefficients as functions of the scattering energy and the Dunkl deformation parameter. In both reflection sectors, the overall evolution of the scattering coefficients remains qualitatively similar. However, the two surfaces become clearly distinguishable in the vicinity of the barrier region, indicating that the Dunkl deformation affects the scattering process differently for $\omega=+1$ and $\omega=-1$. As the scattering energy increases further, the difference between the two surfaces gradually diminishes, suggesting that the influence of the reflection-dependent Dunkl contribution becomes progressively less significant at higher energies.

\subsection{Metastable-State Solutions}

The narrow resonance-like structures observed in the transmission spectrum shown in Fig.~\ref{fig:ScatteringPlots} indicate the existence of metastable (quasibound) states embedded in the scattering continuum. Although these states are not true bound states, they remain temporarily localized inside the sWS barrier before escaping through quantum tunneling. Such states are conveniently described within the Gamow formalism by allowing the energy eigenvalue to become complex,
\begin{equation}
E_m=E_r-iE_i,
\end{equation}
where $E_r$ denotes the resonance energy and $E_i>0$ determines the decay rate of the metastable state.

The corresponding Gamow-state solutions are obtained by analytically continuing the scattering solutions into the complex-energy plane. Consequently, the wave numbers appearing in the scattering solutions become complex, while the hypergeometric parameters remain those defined in Eq.~(\ref{eq:HyperParameters}) with the substitutions $\epsilon\rightarrow\epsilon_m$ and $\nu\rightarrow\nu_m$.

The asymptotic forms of the Gamow wave functions follow directly from the analytically continued scattering solutions. As $x\rightarrow-\infty$,
\begin{equation}
\phi_{m,l}(x)\sim
A_{m,l}e^{\epsilon_m a(x+L)}
+
B_{m,l}e^{-\epsilon_m a(x+L)},
\label{eq:MetastableWaveFuncLeftSideMinusInfty}
\end{equation}
whereas, as $x\rightarrow\infty$,
\begin{equation}
\phi_{m,r}(x)\sim
A_{m,r}e^{-\epsilon_m a(x-L)}
+
B_{m,r}e^{\epsilon_m a(x-L)}.
\label{eq:MetastableWaveFuncRightSidePlusInfty}
\end{equation}

The defining feature of a Gamow state is the absence of incoming waves in the asymptotic regions. Consequently, the amplitudes of the incoming waves vanish, leading to the Siegert boundary conditions
\begin{equation}
A_{m,l}=0,
\qquad
A_{m,r}=0.
\label{eq:GamowBoundaryConditions}
\end{equation}
The resonance energies are determined by imposing the continuity of the wave function and its first derivative at $x=0$,
\begin{equation}
\phi_{m,l}(0)=\phi_{m,r}(0),
\qquad
\phi_{m,l}'(0)=\phi_{m,r}'(0),
\label{eq:MetastableMatchingConditions}
\end{equation}
which yields a homogeneous system of linear equations for the outgoing-wave amplitudes. The resulting nonlinear system is solved numerically in the complex-energy plane to obtain the metastable states of the sWS barrier. The calculated complex resonance energies are listed in Table~\ref{tab:MetastableStates}.

\begin{table}[htb!]
\centering
\caption{Complex resonance energies, decay widths, and corresponding lifetimes for the metastable states of the sWS barrier.}
\label{tab:MetastableStates}
\begin{tabular}{cccc}
\toprule
$\omega$ &
$E_m=E_r-iE_i$ (MeV) &
$\Gamma$ (MeV) &
$\tau$ (s) \\
\midrule
$+1$
&
$94.74415455-2.284758\times10^{-13}i$
&
$4.56952\times10^{-13}$
&
$1.44046\times10^{-9}$
\\
$-1$
&
$94.99600881-2.964205\times10^{-13}i$
&
$5.92841\times10^{-13}$
&
$1.11028\times10^{-9}$
\\
\bottomrule
\end{tabular}
\end{table}

For each resonance listed in Table~\ref{tab:MetastableStates}, the decay width and the corresponding lifetime are obtained from
\begin{equation}
\Gamma=2E_i,
\label{eq:DecayWidth}
\end{equation}
and
\begin{equation}
\tau=\frac{\hbar}{\Gamma}.
\label{eq:Lifetime}
\end{equation}
The extremely small values of $\Gamma$ indicate that both resonances are long-lived metastable states.

Evaluating the scattering coefficients at the resonance energies yields
\begin{equation}
\begin{aligned}
R &=0.004143, \qquad
T =0.995857,
\qquad (\omega=+1),\\
R &=0.004718, \qquad
T =0.995282,
\qquad (\omega=-1).
\end{aligned}
\end{equation}

The resonance energies listed in Table~\ref{tab:MetastableStates} coincide with the narrow transmission peaks displayed in the insets of Fig.~\ref{fig:ScatteringPlots}. The transmission probabilities remain very close to unity, whereas the reflection probabilities are nearly zero. This confirms that the narrow transmission peaks originate from long-lived metastable states supported by the sWS barrier.

\section{Conclusion}\label{Sec:sec5}

In this work, we have presented a unified analytical treatment of the bound, scattering, and Gamow (metastable) resonance states of the one-dimensional symmetric Woods--Saxon (sWS) potential within the framework of Dunkl quantum mechanics. By exploiting the reflection symmetry of the Dunkl Hamiltonian together with a Pekeris-type approximation for the inverse-square interaction generated by the Dunkl deformation, the corresponding Schr\"odinger equation was reduced to the hypergeometric form. This approach enabled us to derive analytical wave functions, bound-state quantization conditions, and scattering amplitudes within a common mathematical framework.

Our analysis shows that the Dunkl deformation produces measurable modifications throughout the entire spectrum of quantum states. In the bound-state regime, the deformation lifts the degeneracy between the two reflection sectors, $\omega=\pm1$, leading to distinct energy spectra and probability densities while preserving the parity of the corresponding wave functions. In the scattering regime, the reflection- and transmission-coefficient profiles exhibit a clear dependence on both the Dunkl deformation parameter and the reflection-sector eigenvalue, with the largest differences occurring near the barrier region. Extending the formalism into the complex-energy plane further allowed us to identify the associated Gamow resonance states. The calculated resonance energies, decay widths, and lifetimes reveal that the Dunkl deformation influences not only the bound-state spectrum and scattering observables but also the stability of the metastable states. The excellent agreement between the complex resonance energies and the narrow transmission peaks provides a consistent physical interpretation of the resonant tunneling mechanism through the sWS barrier.

The present work therefore establishes a unified analytical framework for describing bound states, scattering states, and metastable resonance states in one-dimensional Dunkl quantum mechanics. Owing to its general analytical structure, the proposed formalism can be extended to generalized Woods--Saxon interactions, relativistic wave equations, position-dependent mass systems, and other quantum models incorporating Dunkl-type operators. We expect that the present results will contribute to future investigations of quantum systems with reflection symmetry and deformation-induced interactions.

\section*{Declaration of Competing Interest}
The authors declare that they have no known competing financial interests or personal relationships that could have appeared to influence the work reported in this paper.

\section*{Data Availability}
No data was used for the research described in the article.

\section*{Acknowledgments}
B. C. L. is grateful to the Excellence project FoS UHK 2205/2025-2026 for the financial support.

\bibliographystyle{ieeetr}
\bibliography{Dunklmaster}

\end{document}